\documentclass[lettersize,journal]{IEEEtran}
\usepackage{amsmath,amsfonts}
\usepackage{algorithmic}
\usepackage{algorithm}
\usepackage{array}
\usepackage[caption=false,font=normalsize,labelfont=sf,textfont=sf]{subfig}
\usepackage{textcomp}
\usepackage{stfloats}
\usepackage{url}
\usepackage{verbatim}
\usepackage{graphicx}
\usepackage{cite}
\usepackage{svg}
\begin{document}

\title{
Preserving Real-World Finger Dexterity Using a Lightweight Fingertip Haptic Device for Virtual Dexterous Manipulation
% Preserving Finger Dexterity and Providing Contact Sensation with a Lightweight Fingertip Haptic Device for Enhanced Physics-Based Virtual Manipulation 
% Preserving Real-World Finger Dexterity Using a Lightweight Fingertip Haptic Device for Virtual Dexterous Manipulation
% A lightweight fingernail haptic device preserves finger dexterity for virtual manipulation.
}

%\author{IEEE Publication Technology,~\IEEEmembership{Staff,~IEEE,}
\author{Yunxiu Xu, Siyu Wang, Shoichi Hasegawa
        % \textless-this % stops a space
\thanks{This work was supported by JST, the establishment of university fellowships towards the creation of science and technology innovation, Grant Number JPMJFS2112, and JSPS KAKENHI Grant Number 23H03432.

Y. XU, S. Wang and S. Hasegawa are with the Department of Information
and Communications Engineering, School of Engineering, Tokyo Institute of
Technology, Japan, e-mail: \{yunxiu, siw131, hase\}@haselab.net}}% \textless-this % stops a space
%\thanks{Manuscript received April 19, 2021; revised August 16, 2021.}}

\markboth{IEEE TRANSACTIONS ON HAPTICS,~Vol.~x, No.~x, August~x(Preprint)}%
{Shell \MakeLowercase{\textit{et al.}}: A Sample Article Using IEEEtran.cls for IEEE Journals}

\IEEEpubid{0000--0000/00\$00.00~\copyright~2021 IEEE}

\maketitle

\begin{abstract}
This study presents a lightweight, wearable fingertip haptic device that provides physics-based haptic feedback for dexterous manipulation in virtual environments without hindering real-world interactions. The device's design utilizes thin strings and actuators attached to the fingernails, minimizing the weight (1.76g each finger) while preserving finger flexibility. Multiple types of haptic feedback are simulated by integrating the software with a physics engine. Experiments evaluate the device's performance in pressure perception, slip feedback, and typical dexterous manipulation tasks. and daily operations, while subjective assessments gather user experiences. Results demonstrate that participants can perceive and respond to pressure and vibration feedback. These limited haptic cues are crucial as they significantly enhance efficiency in virtual dexterous manipulation tasks. The device's ability to preserve tactile sensations and minimize hindrance to real-world operations is a key advantage over glove-type haptic devices. This research offers a potential solution for designing haptic interfaces that balance lightweight, haptic feedback for dexterous manipulation and daily wearability.

\end{abstract}

\begin{IEEEkeywords}
Virtual reality, lightweight haptic display, tactile devices, dexterous manipulation, wearable haptics
\end{IEEEkeywords}

\section{Introduction}
%\IEEEPARstart 
Researchers and engineers are working on creating a new world capable of reproducing the multi-sensory experiences found in the real world. In this exploration process, how humans interact with the world through touch has become a focal point of study, particularly in the field of virtual reality(VR). As hands are our primary tools for interacting with the world, developing technologies that simulate the haptic experiences of the hands is one of the core topics in haptics. It can enhance users' abilities to interact with, sense, and manipulate three-dimensional items in virtual worlds. The lack of effective hand haptic feedback can lead to inconsistencies between visual perception and bodily perception in virtual environments. This inconsistency not only reduces the users' sense of immersion and satisfaction but also decreases the accuracy and success rate of operations\cite{macfarlane1999force, mcmahan2011tool, nagano2020tactile}.

A major challenge for haptic devices is making haptic technology more appealing to users in everyday life. The common use of such devices faces two main limitations: 
\begin{figure}[!t]
\centering
\includegraphics[width=2.8in]{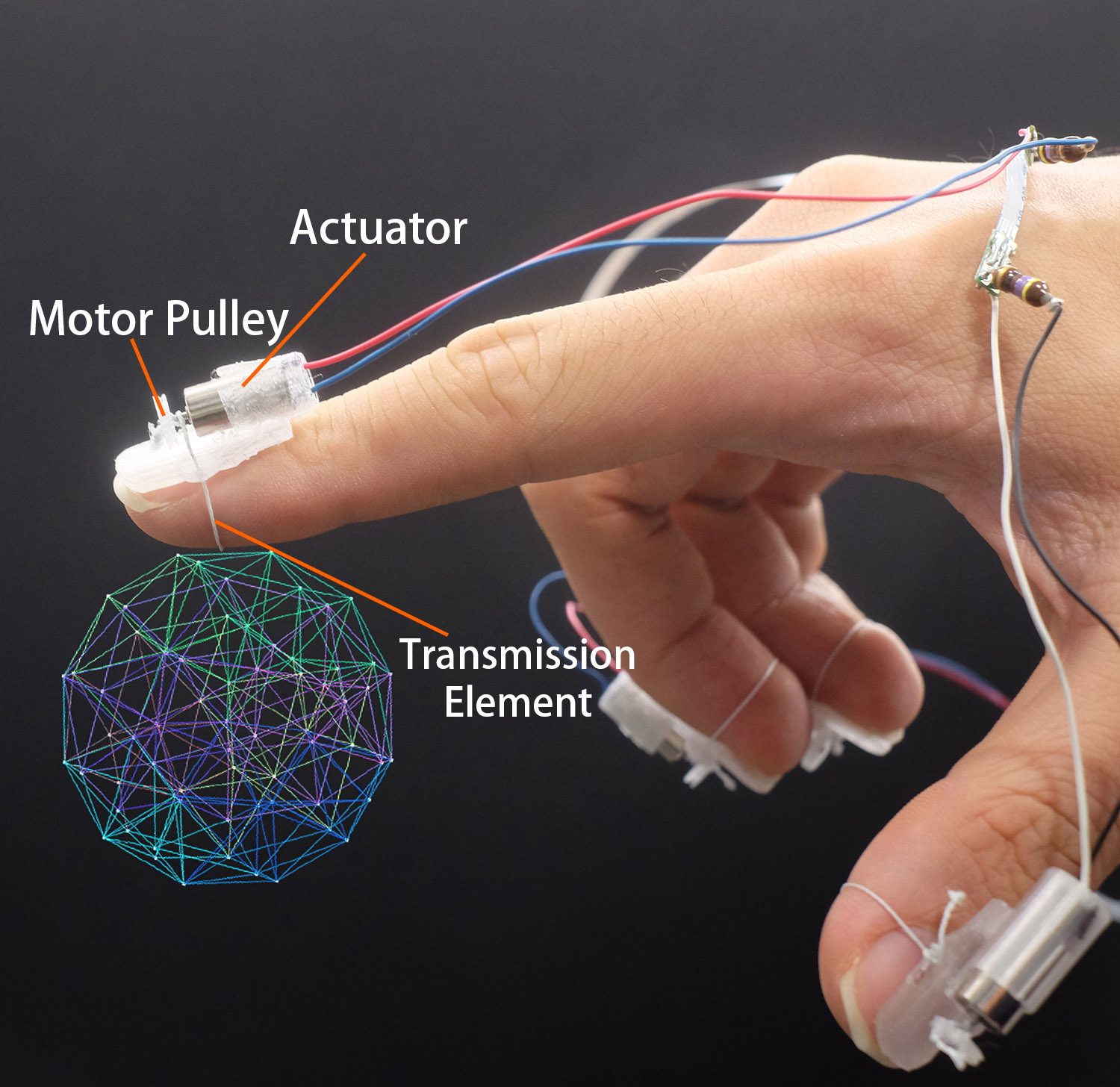}
\caption{The fingertip haptic device is designed to be worn on the fingernail, providing haptic feedback through a thin string driven by a motor. This design keeps the pad of the fingertip unobstructed, allowing the user to feel real objects while receiving haptic sensations. The device weighs 3.57g for the thumb and 1.76g for the other fingers.}
\label{devicePhoto}
\end{figure}
The first limitation is miniaturization and lightness. According to the study by Aoki et al. \cite{aoki2009wearable}, to meet the demands of daily long-time wear, fingertip wearable haptic devices should weigh less than 2.1g, ideally even lighter at 1.4g. Moreover, many studies primarily focus on developing devices capable of simulating specific haptic experiences. While these devices are good at displaying limited particular haptic sensations, adding additional types of haptic feedback while maintaining an acceptable size and weight presents a challenge.
Given the difficulty of replicating all sensations, choosing to preserve the haptic cues that are more critical to virtual interactions becomes important for the device's miniaturization.

\IEEEpubidadjcol

The second limitation is that many haptic devices may hinder users' natural manipulation of physical objects. To encourage the frequent daily use of haptic devices, it is crucial that they minimally impact users' ability to manipulate real objects while providing haptic sensations of virtual objects. Even if it's not possible to fully satisfy comfort requirements, it should at least ensure that users do not tend to frequently remove the device when switching between virtual and real environments. While grip\cite{kim2002tension} or pen-style\cite{silva2009phantom} devices provide accurate manipulation and require minimal time to put on, they do not allow for full finger grasping. Replicating the finger-grasping functionality will require a full-finger device.

This study focuses on providing a solution: developing haptic devices that are both compact and lightweight, do not interfere with daily life, and can effectively provide full-finger physics-based, high-precision haptic feedback for dexterous manipulation in the virtual world. Researchers have developed various wearable haptic devices for virtual and mixed reality, simulating tactile sensations through stimuli like skin deformation, pressure, vibration\cite{yem2017wearable}, and multi-degree-of-freedom force feedback\cite{Prattichizzo2013Towards}. However, these devices often occupy the fingertip area and require further miniaturization and weight reduction.

Some approaches for miniaturization include thin film structures\cite{withana2018tacttoo}, which may sacrifice the texture and friction information of real objects; 
Electroosmotic devices may cover the finger pad \cite{shen2023fluid}, while hydraulic devices require large pumps \cite{feng2017submerged},
and electro-tactile devices \cite{tanaka2023full}, which are compact but struggle to stimulate all mechanoreceptors and simulate friction and pressure.
Also, to avoid hindering finger flexibility, some studies apply feedback to the wrist, finger base\cite{de2018enhancing}, and other locations. However, these areas usually have lower tactile sensitivity compared to the fingertips. Another approach is to remove the actuator when the display haptic is not needed\cite{solazzi2010design}, but it needs time for the actuator to react.

The main challenge is to generate pressure-based haptic without occupying the palmar side of the hand and hindering finger movement. It's also important to avoid generating additional forces outside the expected feedback area. Since most of the receptors are located at the fingertips, we propose a method that uses thin strings as the primary source of fingertip haptic feedback and attaches the actuator to the fingernail. This approach omits the implementation of external forces to control finger movement, but we expect users still to have a response similar to the presence of resistance. Combined with a physics engine, the proposed device can simulate haptic such as pressure, collision, friction, and stick-slip in a virtual environment while still being able to interact with real objects. To maintain the portability of the device, our design does not include the presentation of tangential forces. Therefore, we further investigated the extent to which the tactile feedback provided by the device alone, without tangential forces, facilitates common dexterous manipulation tasks and tested their perception of pressure and their subjective experience with the device.

\begin{figure}[!t]
\centering
\includegraphics[width=3.5in]{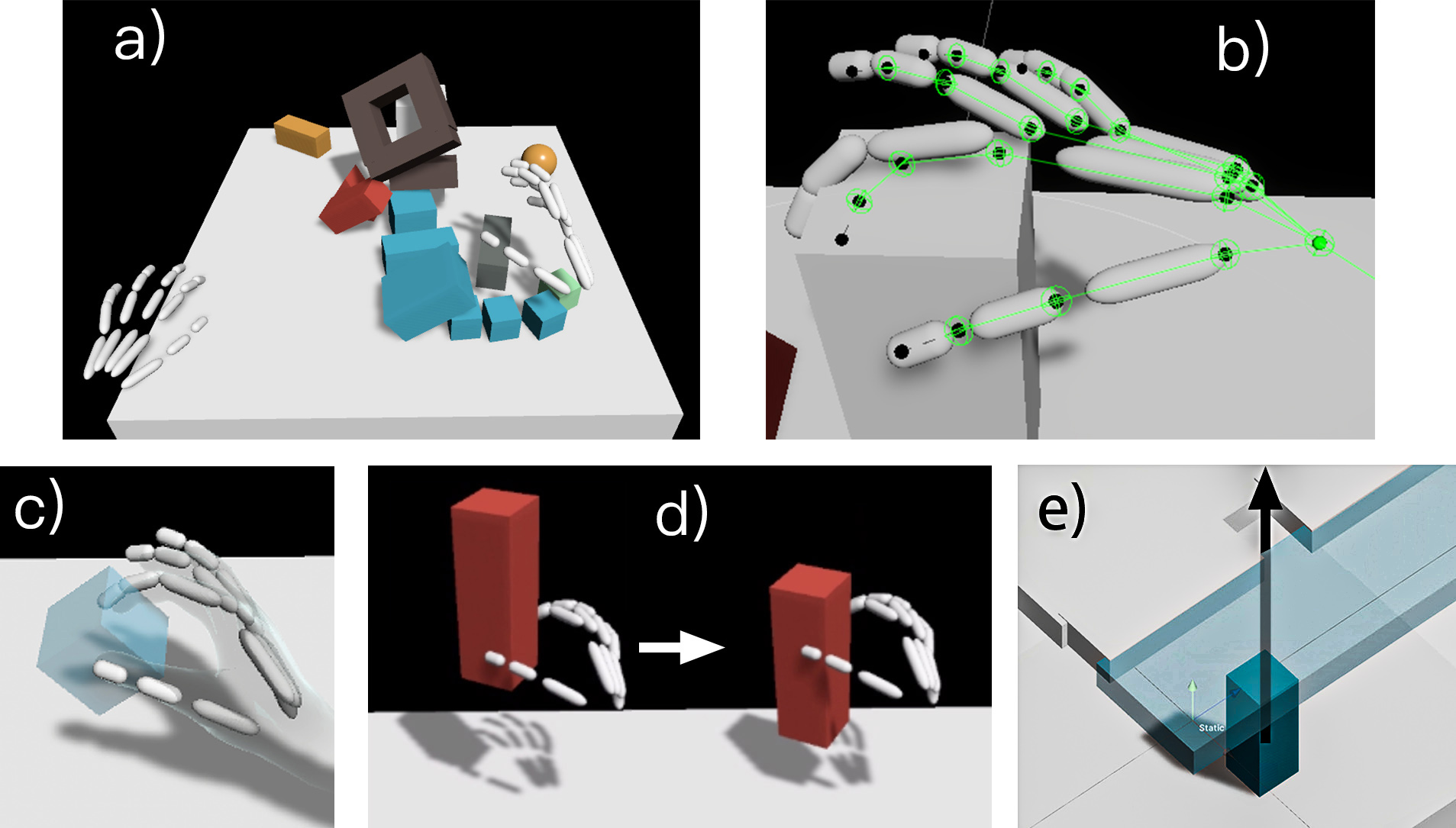}
\caption{a) A sample physics-based dexterous manipulation scene where participants can perform tasks such as stacking objects, placing objects in holes, and sliding objects. b) The virtual hand is composed of virtual couplings, with the force output proportional to the penetration depth of the real hand's position(in green). The measured phalanges and phalanges in the physics engine are connected by virtual springs. Also, adjacent phalanges are connected by ball joints. c) In the fingertip pressure perception experiment, participants were asked to grasp an object and find the smallest grasp force that would not cause the object to drop. d) The "Re-Grasp" experiment. participants were asked to slide an object and re-grasp it before it fell to the ground. e) Peg-in-hole, a typical dexterous manipulation task, requires participants to insert the object into a designated hole.}
\label{UnityScene}
\end{figure}

\section{Related Works}
One of our primary goals is to achieve dexterous manipulation in the virtual world. Han et al.\cite{han1998dextrous} defined dexterous manipulation, considering the workspace limitations of the fingers, using rolling and finger gaiting to adjust the grasping posture, and changing the contact points through finger rolling and sliding. Bicchi et al.\cite{Bicchi2000Hands} believe that controlled slippage and rolling contact can improve the sensitivity of robotic manipulation. To achieve this, it is necessary to predict the occurrence of slippage and detect changes in the contact point position on the finger surface. The process of dexterous manipulation may need to obtain the motion state of the object, the contact state between the object and the fingers, and the friction characteristics. Some studies have explored the control principles of the human brain for dexterous manipulation. Previous research \cite{Johansson2009Coding} \cite{Johansson1987Signals} believes that when an object undergoes unexpected acceleration due to instantaneous sliding between the fingertips and the object, automatic adjustment of the grasping force can be observed. Flanagan et al.\cite{Flanagan1993Coupling} shows that human grip force is finely adjusted with changes in load force, indicating a coupling between grip force and load force. Wiertlewski et al.\cite{Wiertlewski2013Slip} found that the central nervous system can perceive the movement state of an object through vibrations between the skin and the object's surface and adjust the grasping force to terminate sliding. To create hand haptic, we need to stimulate mechanoreceptors critical for manipulation\cite{johansson1983tactile, johnson2001roles}. If we can provide the necessary information based on physical simulations, improving the experience of dexterous manipulation of virtual objects is possible.

Next, we discuss the research on devices for reproducing haptic. Haptic displays can be considered devices that generate mechanical impedances \cite{Lin2008Haptic}. Currently, there are various haptic devices aimed at applying impedances to the fingers to present tactile sensations, such as haptic gloves \cite{Qi2023HaptGlove}, hand-held devices \cite{choi2018claw,Benko2016Normaltouch}, and devices that control finger activity through locking mechanisms \cite{Choi2016Wolverine,Hinchet2018Dextres, Fang2020Wireality} uses shoulder-mounted cable drives to lock the hand joints. These devices control hand movements to display pressure sensations, requiring a stable support point like the wrist, arm, or shoulder to withstand the reaction force. Therefore, the overall volume of these solutions is difficult to reduce, and they decrease hand comfort and freedom.

However, many studies focus on implementing forces only at the fingertips, without applying impedances to the entire finger to control finger movement. These are finger-worn haptic devices. Schorr et al.\cite{schorr2017fingertip} developed a device that can generate lateral skin deformation. There are also some researches that use a dual-motor mechanism to reproduce the deformation of the finger pad\cite{Minamizawa2007Gravity}\cite{tsetserukou2014linktouch}. Prattichizzo et al.\cite{Prattichizzo2013Towards} proposed a 3-degree-of-freedom wearable fingertip tactile device that can achieve tangential force and pressure feedback. Girard et al.\cite{girard2016haptip} display two-degree-of-freedom(DOF) shear force feedback on the fingertip, and Leonardis et al.\cite{leonardis20163} achieves deformation of the fingertip in three 3DoF. Yem et al.\cite{yem2017wearable} created a device that can present high-frequency vibrations and skin deformation. The above studies aim to achieve tangential force or shear force while allowing relatively free finger movement, but they require a certain volume of mechanical structure to be placed on the fingertip. 
More importantly, when wearing VR devices, we still need to perform real-world operations frequently, such as using a real keyboard, which becomes very difficult if fingertip devices hinder natural sensations and finger movements.
Additionally, some devices also require the integration of finger posture sensors or markers, which further increases the size of the devices.

Furthermore, some researchers choose not to implement tangential force or shear force to reduce the device volume. For example, Vechev et al. \cite{Vechev2019Tactiles} designed a small actuator array with a smaller individual volume ($1\,\text{cm}^3$) but still needs to be placed on the palmar side. Therefore, some studies attempt to "clear" the palmar side to preserve the tactile sensation of the fingertips. One method is to directly transfer the tactile actuator to other parts of the hand, such as placing it on the fingernail \cite{Ando2002SmartFinger,Preechayasomboon2021Haplets,Rekimoto2009SenseableRays}. Other studies relocate hand tactile sensations to the wrist \cite{Pezent2019Tasbi,Sarac2022Effects,Moriyama2022Wearable,Casini2015Design} or install the device on the proximal phalanx \cite{pacchierotti2016hring}. Maeda et al.\cite{Maeda2022Fingeret} place two motors on both sides of the finger and compress the skin by rolling to free the finger pad. Although these methods can preserve the skin's tactile sensation of the fingertips and are easy to wear, they cannot create pressure stimulation at the fingertips, which is common when grasping objects.

One approach is to move the palmar side components to other positions when not needed, not contacting the fingers in the non-contact state\cite{solazzi2010design,teng2021touch,kovacs2020haptic, mercado2021haptics}. The disadvantages include the inability to touch real and virtual objects simultaneously, the longer response time required by the mechanical structure and the actuator, and sometimes being too large to be fixed to fingers or hands, these additional structures also increase the volume. Additionally, Tsai et al.\cite{tsai2022fingerx} use extendable and retractable supports on the fingers to render the tactile shape of virtual objects. Due to its size, this device limits dexterous interactions with virtual and real objects.

Another method is to completely remove the actuator from the human body, such as ultrasound \cite{Matsubayashi2019Direct}, air \cite{Gupta2013Airwave}, and synthetic jets \cite{Shultz2022LRAir}. These methods provide contactless tactile feedback. However, they often require a relatively unobstructed transmission, and if the device is not "encounter-type"\cite{mercado2021haptics}, the effective range is usually limited.

The proposed device keeps the actuator on the palmar side, an approach that has also been explored in several studies. Electrostimulation does not require mechanical structures and is suitable for achieving thin tactile sensations. 
Tanaka et al. \cite{tanaka2023full} used electrostimulation to achieve tactile sensations on the palmar side without placing electrodes on this side. However, it cannot achieve continuous pressure feedback. Moreover, the electrostimulation method cannot selectively stimulate Pacinian corpuscles \cite{kajimoto2004electro}, which may lead to difficulty in achieving tactile sensations above 100Hz, which are closely related to sliding. Withana et al.\cite{withana2018tacttoo} present a thin-film tactile actuator that can achieve almost bare-skin tactile sensitivity, but this method still affects users' perception of texture. Furthermore, the contact area can also be reduced. Han et al.\cite{han2018hydroring} use fluid to provide pressure on the finger pad, but it is not easy to generate high-frequency tactile signals. The actuator can also be placed on the fingernail side to stretch light and small objects. Aoki et al.\cite{aoki2009wearable} placed the actuator on the fingernail and used a thin string, which can generate pressure at the finger pad. Our work is also based on this research, but this research has not yet achieved full installation on all five fingers and hasn't achieved physics-based tactile rendering. In addition, the static tension was too small to feel, and dynamic modulation was needed. 
Based on our previous work\cite{xu2023realistic}, this paper removes the ring structure that occupies the palmar side and conducts user studies and performance evaluations.

\begin{figure}[!t]
\centering
\includegraphics[width=3.5in]{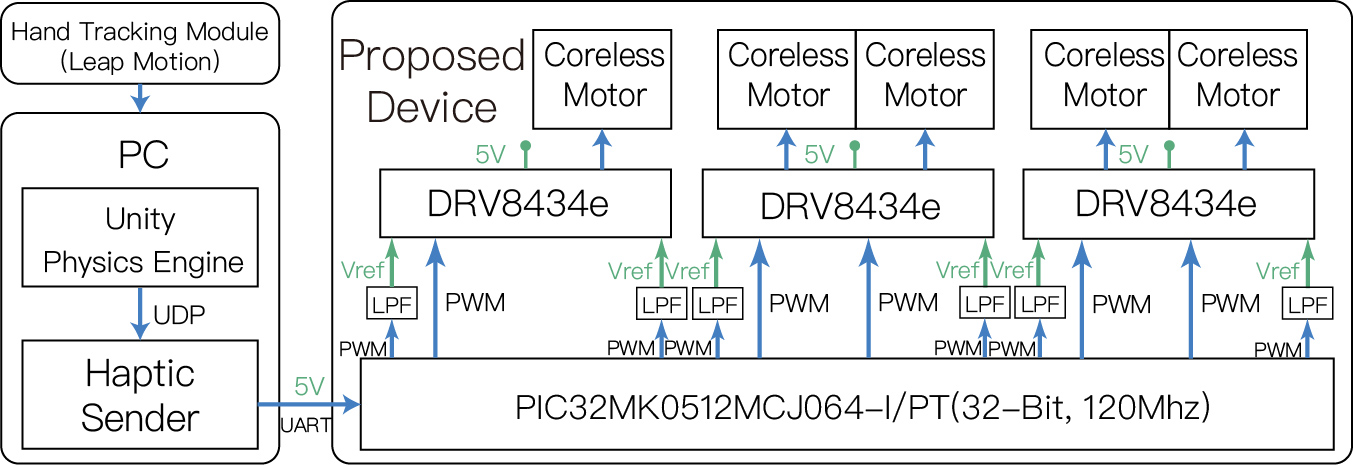}
\caption{The schematic diagram illustrates the information flow and main components of the system.}
\label{systemFigure}
\end{figure}

\section{System Overview}

\begin{figure}[!t]
\centering
\includegraphics[width=3.0in]{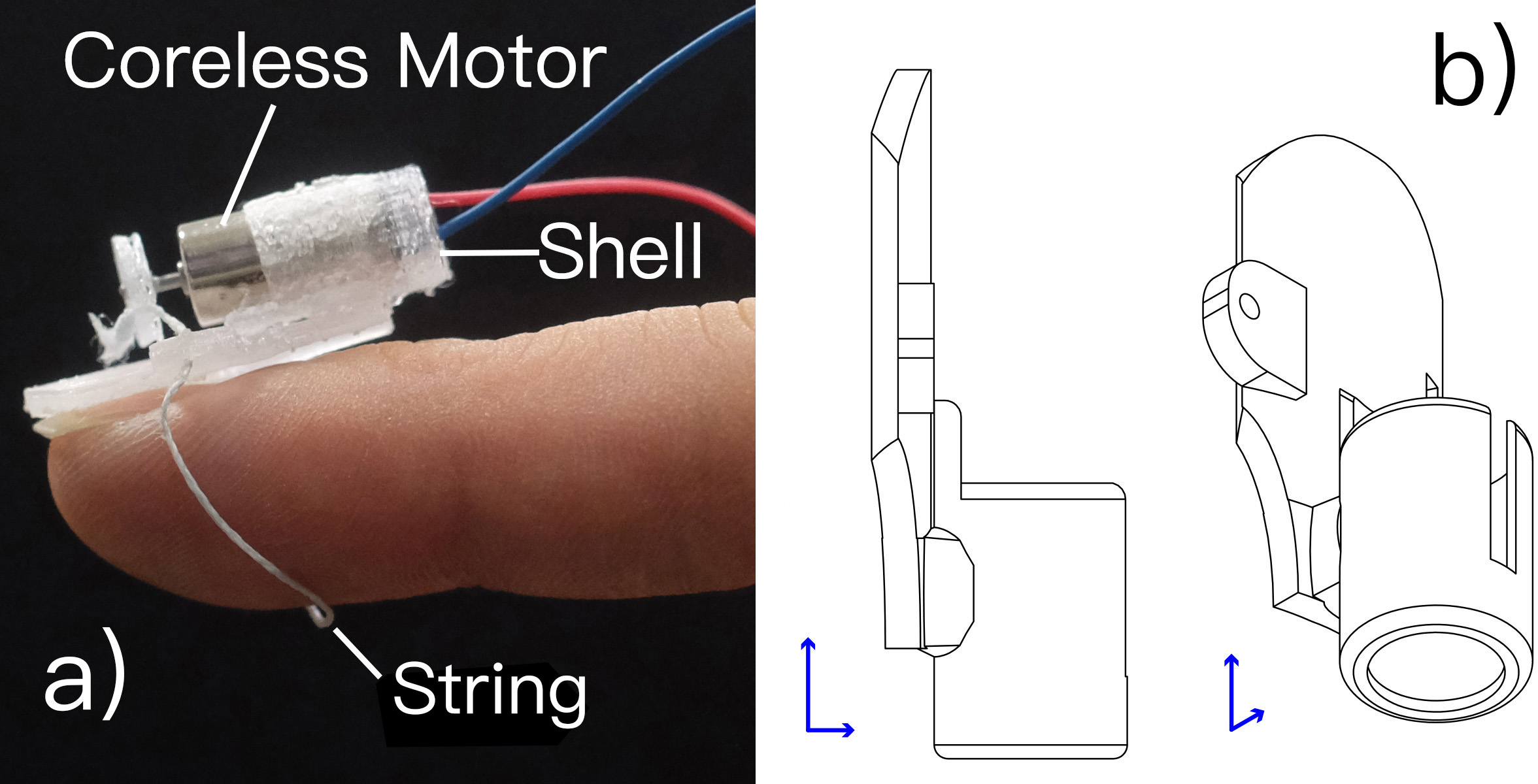}
\caption{The hardware components are mounted on the fingertip, with the shell adhered to the fingernail for stability. The device has a weight of 3.57 grams to the thumb's fingertip and 1.76 grams to each of the other fingertips.}
\label{smallDevicePhoto}
\end{figure}

\subsection{Implementation}
The system overview is shown in Figure \ref{systemFigure}. 
\subsubsection{Hardware}
Our hardware was composed of a microcontroller, motor driver, and actuator. We chose the PIC32MK0512MCJ064-I/PT for the microcontroller, which receives serial port signals from the PC containing haptic information. This controller is capable of outputting up to 9 channels of PWM signals, which can meet our requirements to control multiple motors. We employed the driver with the current controller DRV8434e to control the output current of each motor. Then, we can convert the haptic signals into currents to control the motors.

Regarding the actuator and shell, we chose coreless motors with a size of 6mm in diameter and 12mm in length (7mm in diameter and 16mm in length for the thumb). This type of motor is light and cheap.

To attach the device to the fingernail, we used acrylic-modified silicone adhesive (CEMEDINE BBX909), which can be pasted and peeled off, maintaining adhesiveness after multiple uses. We are using PE string with a diameter of 0.260mm.
About the design of the axle for winding the string, assuming the motor's torque is T and the radius of the motor's axle is r, the pull force F on the finger can be approximated as:
\begin{align}
F = \frac{T}{r}
\end{align}
Therefore, the radius of the motor's axle should be as small as possible. After experimenting with various axle designs, we ultimately chose to use the motor's own axle due to its thin diameter of 0.8mm. The length of the string is also crucial. According to the Capstan equation: 
\begin{align}
T_{1} = T_{0} e^{\mu\theta}
\end{align}
$T_{1}$ is the tension in the rope as it leaves the cylinder.
$T_{0}$ is the tension in the rope as it contacts the cylinder. $\theta$ is the angle, in radians, that the string wraps around the cylinder.$\mu$ is the coefficient of friction between the rope and the cylinder. 
The equation shows that minimizing the rope length around the motor’s axle reduces resistance. However, overly short ropes maintain maximum tension and limit pressure. We found that the optimal rope length just touches the finger surface without applying pressure when the motor is off. The fingertip hardware is shown in Figure \ref{smallDevicePhoto}.

\subsubsection{Software}
For the software component, we utilized Unity to create virtual scenes and visual effects. Regarding the physics engine and high-quality haptic rendering, we employed Springhead\footnote{https://springhead.info/} as the physics engine. Compared to Unity, Springhead can obtain richer information, such as contact force between rigid bodies and dynamic or static friction states. 
We used virtual couplings to build hands, which helps maintain system stability and ensures that the calculated feedback force is proportional to the penetration depth
, as shown in Figure \ref{UnityScene}b. After constructing the virtual scene, we obtained the penetration force of the finger's rigid body through the physics engine, corresponding to the pressure generated by the haptic device. The instant when the penetration force changes from absent to present correspond to the collision vibration, and the moment when the friction force transitions from static to dynamic friction corresponds to the stick slip.

We need a high update rate to generate high-quality, stable multi-channel haptic signals. To avoid impacting Unity's performance, we set it only to transmit changing components of the haptic signal (e.g., the moment of collision). The generation of haptic signals is handled by a separate independent program.

\begin{figure}[!t]
\centering
\includegraphics[width=3in]{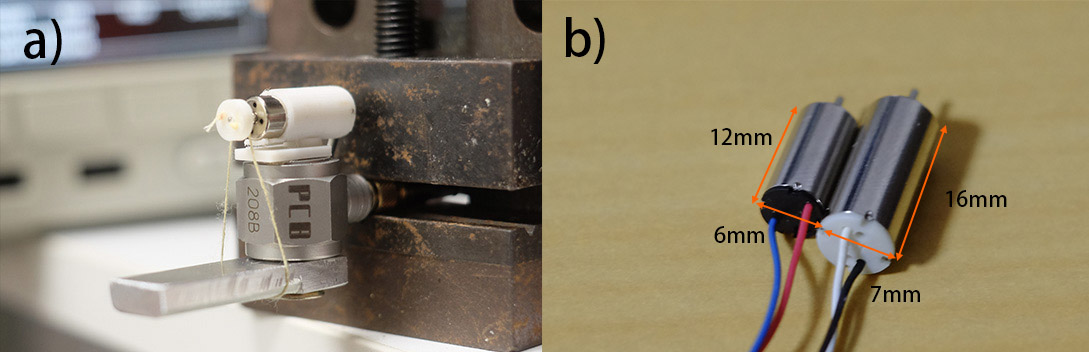}
\caption{a): Measuring the pulling force of the motor using a force sensor, the clamp fixed the system to decrease vibration. b): Comparing the shapes of coreless motors for the thumb and other fingers; 612 denotes a diameter of 6mm and a length of 12mm, while 716 indicates a diameter of 7mm and a length of 16mm.}
\label{mesaureSet}
\end{figure}

\subsection{Hardware evaluation}

\begin{figure}[!t]
\centering
\includegraphics[width=2.5in]{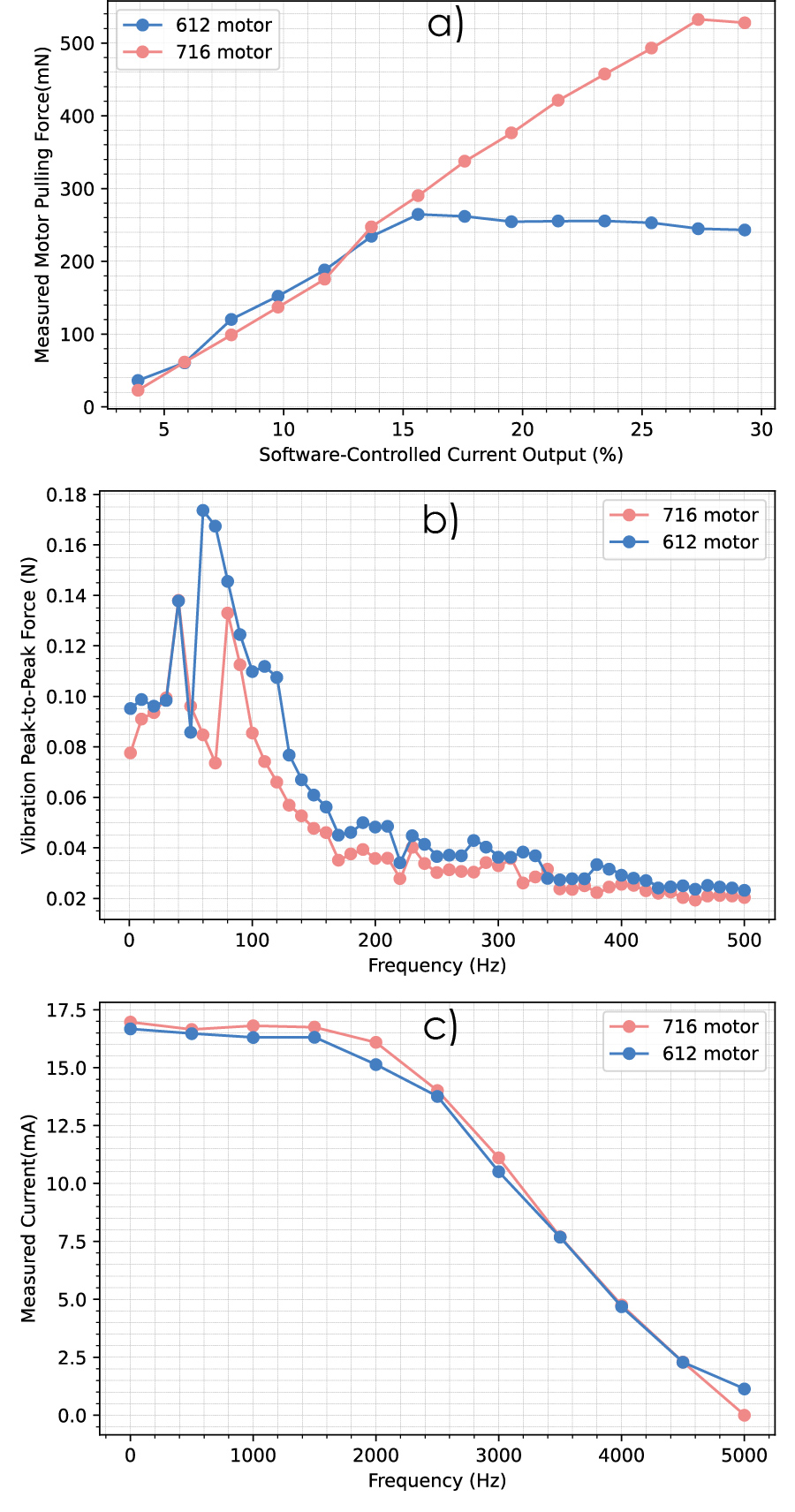}
\caption{a): Motor pulling force response to software control. b): Frequency response of motor vibration intensity. It can be observed that there are strong resonance points within the range of 80 to 100Hz. c): Frequency response of motor current, the current can almost remain stable below 1500Hz}
\label{MotorForce}
\end{figure}

Our focus on the device's hardware performance is the force and the vibration the motor can provide. These two metrics are crucial in determining the pressure and vibration feedback level we can deliver to the fingertips. Our system used 2 types of motors shown in Figure \ref{mesaureSet} b(larger one for thumb).

We first measured the motor's output force to test the feasibility of achieving linear force control and to determine its output range. 
We used a PCB Piezotronics 208B force sensor to measure the force, with the measuring system shown in Figure \ref{mesaureSet}a. 
We designed this measuring structure to simulate the proposed device, enabling sensors to measure the pressure exerted by the string.

The results are shown in Figure \ref{MotorForce}a. The pulling force for the 612 motor, after about 15\% of the current output, basically remains stable, reaching its saturation point of approximately 260mN. Also, the 716 motor exhibits a more significant increase in pulling force. The maximum pulling force is around 540mN. The current and pulling force showed a generally linear relationship. According to a study by Lederman et al.\cite{lederman1999sensing}, the average human finger skin perception threshold is 0.031N. Therefore, we need to output at least 4\% current for it to be perceptible to users.

% \begin{figure}[!t]
% \centering
% \includegraphics[width=1.5in]{img/measureSensor.jpg}
% \caption{Measure frequency response}
% \label{measureSensor}
% \end{figure}

% \begin{figure}[!t]
% \centering
% \includegraphics[width=1.5in]{img/2motor.jpg}
% \caption{This is the caption for one fig.}
% \label{2motor}
% \end{figure}

Next, we assessed the frequency response of the device's vibration intensity, inputting waveforms of various frequencies and measuring the peak-to-peak value of the vibration intensity while keeping the current constant. Additionally, it was necessary to understand the changes in current by different frequencies. We achieved this by connecting a 1 $\Omega$ cement resistor in parallel 
to obtain the peak-to-peak current values. The results are shown in Figure \ref{MotorForce} b and Figure \ref{MotorForce} c. The results indicate that both types of motors have distinct resonance within the 80 to 100 Hz range. Furthermore, the current response of both motors decreases with increasing frequency and remains nearly stable below 1500 Hz.

% \begin{figure}[!t]
% \centering
% \includegraphics[width=3in]{img/MotorCurrent.jpg}
% \caption{The measured current of the motor under the software and the motor driver's current control function. It can be observed that the relationship is essentially linear"}
% \label{MotorCurrent}
% \end{figure}

\section{Fingertip Pressure Perception Experiment}
\label{sec:presureExperiment}

Compared to similar devices, the unique feature of the proposed device is its ability to create pressure on the fingertips. However, due to our aim to design a slim device, the force it can exert is relatively mild. When the user's fingers grasp a virtual object, the proposed device cannot apply a force to stop the fingers' movement, meaning it cannot provide a kinesthetic constraint. This is a key difference from many other devices. Consequently, it is crucial to assess user perception to determine whether they can truly feel the pressure and if this pressure can influence user behavior, resulting in reduced pinch distance.

\subsection{Methods}
\subsubsection{Participants}
A total of 12 participants, 8 male, and 4 female) participated in the experiment after giving informed consent. The participants were between the ages of 23 and 32 and were one left-hand dominant. No special treatment for the left-hand dominant participants. The experiment procedure was approved by the Tokyo Institute of Technology Review Board.

\subsubsection{System Set-Up}
\begin{figure}[!t]
\centering
\includegraphics[width=2.7in]{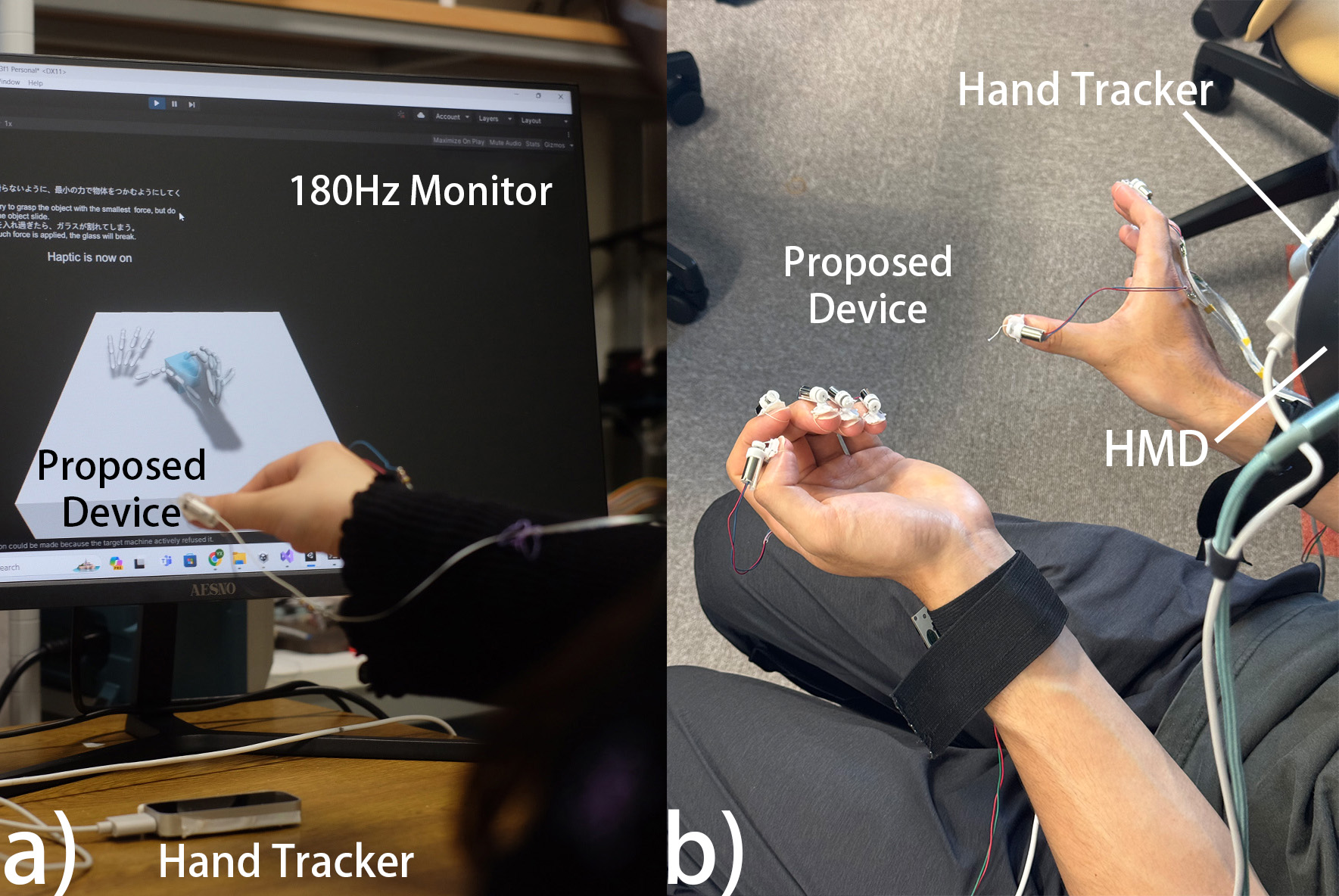}
\caption{a) In the experimental setup for the fingertip pressure perception experiment, participants were instructed to wear the proposed device and use a Leap Motion controller as an input device to complete the experiment. b) The experimental setup for HMD.}
\label{manipulationPhoto}
\end{figure}

The user wore the proposed device on their right hand and used a Leap Motion Controller for hand recognition. Due to the high refresh rate required for the following experiment, a monitor with a 180Hz refresh rate was used as the display device instead of a Head Mounted Display (HMD). The experiment was conducted within a virtual object manipulation environment (Figure \ref{manipulationPhoto} a), with the same scale as the real world. Users interacted with virtual objects using a virtual rigid body hand. 
The virtual scene is shown in Figure \ref{UnityScene} c.
We used a cube with a length of 5 cm and a density of 1.00 for manipulation. We assigned the cube a static friction coefficient of 4.00 and a dynamic friction coefficient of 3.00. This large friction was chosen to prevent the cube from sliding, allowing the experiment to focus on pressure feedback without the interference of sliding sensations. The physics engine's time step was 0.01s.
Throughout the experiment, real-time data were collected on the virtual object's height and the sum force of the user's thumb's rigid body.

% \begin{figure}[!t]
% \centering
% \includegraphics[width=1.5in]{img/222.png}
% \caption{Fingertip pressure perception experiment.}
% \label{222}
% \end{figure}

\subsubsection{Experimental Procedure}

During the experiment, participants wore the proposed device and were instructed to grasp the cube in the virtual environment. They received a prompt: "Please try to grasp the object with the smallest force, but do not let the object slide or drop. If too much force is applied, the glass will break." Before the experiment, participants had time for free play to become familiar with the environment. The virtual cube is transparent, and if the user applies a force greater than 15N, the cube's position resets, indicating that the "glass" has "broken." It was repeatedly emphasized to use the minimal force possible for grasping. The formal experiment begins once the user believes they have achieved the minimum grasping force.

In the formal experiment, the glass never breaks to reduce the number of attempts and expedite the process. However, this information is kept confidential from the participants. Also, users were informed of the procedure before each attempt to prevent forgetting to use minimal force. When users believe they have reached the minimum force, the trial is finished. To investigate if haptic feedback affects pinch distance when grasping objects, we faced an issue where the index finger might be obscured in the tracking camera view. So, we recorded the thumb tip's normal force on the object in the physics engine as a substitute for pinch distance.

The experiment had three conditions: "no haptic", "contact vibration" and "pressure" with the order counterbalanced across participants. Each completed 3 rounds of cyclic trials for every condition. "Contact vibration" refers to a damped sinusoid signal generated during virtual contact between a finger and any rigid body, proportional to the relative velocity between the fingertip and the object, providing a cue for contact start. "Pressure feedback" is delivered through the thin string exerting force on the fingertips.

\subsection{Results}
\begin{figure}[!t]
\centering
\includegraphics[width=2.2in]{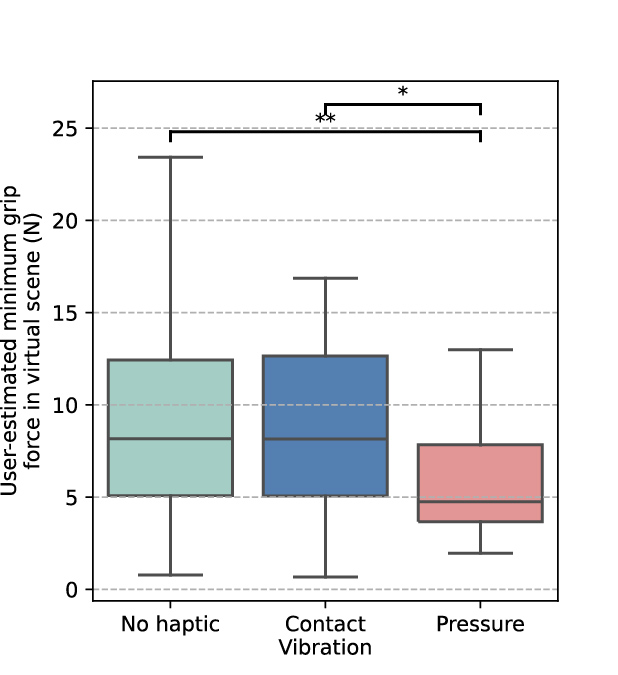}
\caption{The result of participants completed the grasping task of virtual objects under three different feedback conditions; participants were required to grip objects made of glass with the minimum force, as we were told excessive force would result in the object break(***p \textless 0.001, *p \textless 0.01, *p \textless 0.05).}
\label{minPressure1}
\end{figure}

The user-estimated minimum grip force under three different conditions is shown in Figure \ref{minPressure1}. 4 data points were eliminated because they used the middle phalange to grasp objects, resulting in no pressure data detected at the fingertip. Summary statistics revealed that the median values for conditions "no haptic", "contact vibration", and "pressure" were 8.166N (SD=5.881, SEM=0.994), 8.158N (SD=4.725, SEM=0.788), and 4.743N (SD=2.966, SEM=0.516), respectively.
The Shapiro-Wilk test showed that the data for conditions "no haptic" (W=0.958, p\textgreater0.05) and "contact vibration" (W=0.958, p\textgreater0.05) were normally distributed, while the data for condition "pressure" (W=0.929, p\textless0.05) deviated from normality. Levene's test was performed to evaluate homoscedasticity, which indicated unequal variances among the conditions (W=6.085, p\textless0.01). Given these results, we used a non-parametric test to compare the conditions. The Kruskal-Wallis test results showed a significant difference among the conditions (H(2)=8.796, p\textless0.05).

To investigate the pairwise differences between conditions, Mann-Whitney U post-hoc tests with Bonferroni correction were conducted. The post-hoc tests revealed no significant difference between conditions "no haptic" and "contact vibration" (p=1.000, Hedges' g=-0.154). However, significant differences were found between conditions "contact vibration" and "pressure" (p\textless0.05, Hedges' g=0.698) and between conditions "no haptic" and "pressure" (p\textless0.01, Hedges' g=0.765).

% \begin{figure}[!t]
% \centering
% \includegraphics[width=3.2in]{img/minPressure0.jpg}
% \caption{The minimum pressure is perceived to be able to grasp an object of each user. Condition 0: No haptic sensation Condition 1: Collision haptic sensation Condition 2: Pressure haptic sensation}
% \label{minPressure0}
% \end{figure}

 \subsection{Discussion}

The results indicate that the proposed device's pressure feedback can change the grasping task's behavior, suggesting that pressure feedback alone can enable users to find a relatively smaller grasping force. This outcome implies that a device applying pressure to the fingertips can provide haptic feedback of the contact force, which helps in perceiving the grasping state of the object. This could be because continuous pressure feedback stimulates SA1 receptors, and when the fingertips receive pressure stimulation, the brain naturally perceives that the fingers are touching and grasping a real object, even though finger movement is not actually obstructed. Moreover, applying a varying pressure gradient to the fingertips can simulate tactile changes during grasping. For example, gradually increasing pressure can represent the fingers gradually tightening their grip on the object.

However, There was no significant difference in "contact vibration" compared to no haptic feedback. The conclusion is similar to previous research\cite{aoki2009wearable}. The condition "contact vibration" can provide information about the start of contact but does not convey continuous information about the applied force. "Contact vibration" alone may not be sufficient to alter users' behavior in grasping tasks.

\section{Slip Vibrotactile Feedback Experiment}

In our previous research\cite{xuOptimizing}, we observed that participants could grasp an object with their fingers, gently release it to allow sliding, and then immediately grasp it again using only vibration feedback, even in the absence of tangential force. This is a common and challenging dexterous manipulation task. Our prior work also noted that there was no increase in reaction delay compared to situations where tangential force is present. The device proposed in the current paper cannot display tangential force. Therefore, we will investigate whether vibration or pressure alone, generated by the proposed device, can significantly improve users' reaction speed when sliding objects.

\subsection{Methods}

\subsubsection{Participants}
Same as the fingertip pressure perception experiment(section \ref{sec:presureExperiment}).

\subsubsection{System Set-Up}
The hardware setup is the same as the fingertip pressure perception Experiment(section \ref{sec:presureExperiment}). In the experimental scene, we used a virtual cube of 15cm in length, 5cm in width, and 5cm in height. The static friction coefficient was set to 0.15, the density is 1, and the dynamic friction coefficient was set to 0.10 for easy sliding, as shown in Figure \ref{UnityScene}d.

\subsubsection{Experimental Procedure}
Before the formal experiment, participants started pre-training. They could repeatedly practice the task with visual feedback until confident enough to perform it independently. Participants had to lift and maintain their grasp on an object. After steadily grasping the object for a period, they had to control the object to slide between virtual fingers and immediately attempt to re-grasp it to prevent dropping. However, participants were prohibited from quickly releasing and immediately pinching their fingers without feedback to complete the action, as it violates the requirement to keep the object sliding. The experiment had 4 conditions: "No haptic", "pressure,", "pressure and vibration", and "vibration". Pressure feedback involves applying normal force to the fingertips through strings. Vibration feedback occurs when virtual fingers slide on the object's surface and during stick-slip. Condition order was counterbalanced, with each condition attempted five times. If the "re-grasp" was unsuccessful (object not re-grasped after sliding or falling), the result was still recorded, but the fingers' grasping reflex had to be observed. If no reflex was observed, the attempt was retried.

The moment the object started sliding was defined as T2, and the moment the fingers exhibited a grasping response was T1. We recorded T1-T2 for each attempt and considered it the response latency.

\begin{figure}[!t]
\centering
\includegraphics[width=3.0in]{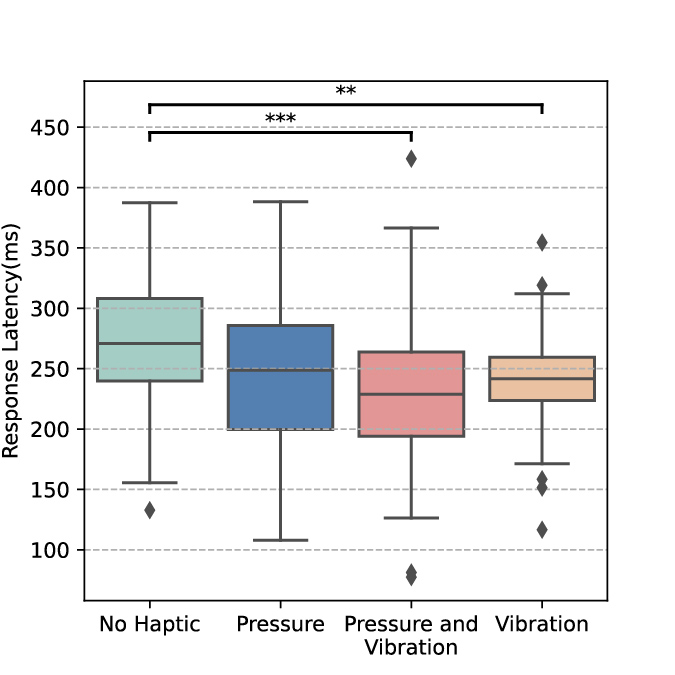}
\caption{Response latency under 4 different haptic feedback conditions in the "re-grasp" task. The moment when the object started to slide was defined as T2, and the moment when the fingers exhibited a grasping response was defined as T1. The latency is T1-T2. }
\label{slidingSuccessRate}
\end{figure}

\subsection{Results}
Figure \ref{slidingSuccessRate} displays the experimental results. 7 data were eliminated from the analysis because the speed of finger opening was too fast, and there was a lack of sliding friction(which wasn't detected when experiencing). The median values for conditions "no haptic", "pressure", "pressure and vibration", and "vibration" were 270.88ms (SD=52.910, SEM=6.831), 248.58ms (SD=60.206, SEM=7.838), 228.77ms (SD=58.512, SEM=7.618), and 241.72ms (SD=41.774, SEM=5.439), respectively. The Shapiro-Wilk test showed that the data for all conditions "no haptic" (W=0.991, p\textgreater0.05), "pressure" (W=0.986, p\textgreater0.05), "pressure and vibration" (W=0.961, p\textgreater0.05), and "vibration" (W=0.970, p\textgreater0.05) were normally distributed. Levene's test was performed to evaluate homoscedasticity, which indicated equal variances among the conditions (W=2.356, p\textgreater0.05).

Given that the data were normally distributed and had equal variances, a parametric one-way ANOVA was used. The results showed a significant difference among the conditions (F(3,233)=6.722, p\textless0.001). Tukey's HSD post-hoc tests were conducted to investigate the pairwise differences between conditions. 
Significant differences were found between conditions "no haptic" and "pressure and vibration" (p\textless0.001) and between conditions "no haptic" and "vibration" (p\textless0.01).

\subsection{Discussion}
Vibration can significantly reduce participants' response time when re-grasping a sliding object, this conclusion aligns with our previous research\cite{xuOptimizing}. However, pressure feedback only is insufficient to improve users' reaction speed in this specific task. This could be due to the pressure feedback primarily providing information about the normal force rather than the tangential force associated with sliding. Additionally, it may be because friction force was not used for the feedback, which may be altered by changing the haptic rendering design. During the experiment, participants often say, "The object feels completely smooth, and it's difficult to perceive the sliding" when switching to the pressure-only mode. Despite the lower minimum value in the "pressure" condition compared to "no haptic," most participants couldn't use pressure feedback to improve reaction time in predicting when the object would fall. Moreover, the insignificant differences among "pressure," "pressure and vibration," and "vibration" conditions imply that combining pressure and vibration feedback may not considerably enhance reaction speed.

Compared to our previous study \cite{xuOptimizing}, the current study shows longer reaction times, likely due to increased latency from the hand tracker and software. The reaction time difference between no haptic and vibration conditions was around 90ms in the previous study, but only around 35ms in the current one. We speculate this could be due to two reasons: 1. The previous study required successful grasping, collecting only the fastest reaction data, while the current experiment only observed pinch reflex without requiring successful manipulation. 2. The small device's weak vibrations may be difficult to perceive during the low-speed initial sliding phase, further increasing reaction delays. Additionally, previous studies involved grasping real objects(the handle of the device), while the current experiment requires finger movement to decrease distance, potentially causing differences. We selected data from four participants who participated in both experiments under vibration-only conditions. 
\section{Evaluation of Haptic Interaction and Wearability}
\begin{figure}[!t]
\centering
\includegraphics[width=2.7in]{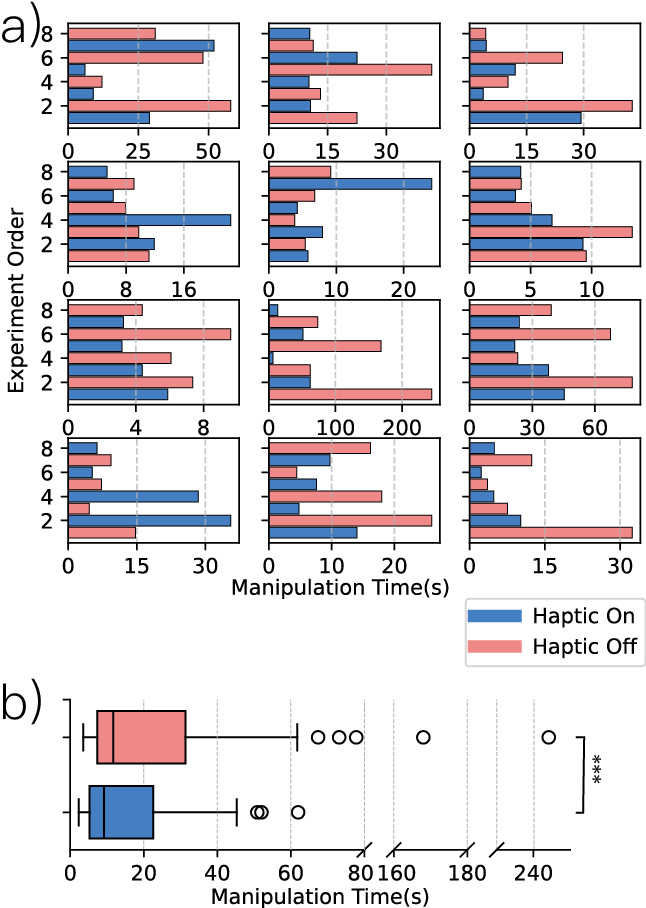}
\caption{Manipulation time data for the peg-in-hole task.
a) Each subplot represents data from 8 trials of a single participant, with the horizontal axis showing the manipulation time and the vertical axis indicating the trial order (from the first to the eighth attempt). The trials alternated between haptic feedback being on or off, with half of the participants starting with haptic feedback on and the other half starting with it off.
b) A box plot of the manipulation time data for all participants.}
\label{peg}
\end{figure}

We will test whether the proposed device improves dexterous manipulation efficiency and understand users' subjective experiences with it. We designed an evaluation to gain insights into participants' thoughts and experiences, focusing on: 1. Manipulating virtual objects: Participants reported on the realism of haptic feedback, the accuracy of force feedback, and the smoothness of operation, along with their operation times. 2. Comparison with the real world: Participants compared virtual grasping using the haptic device to real-world sensations. 3. Device-wearing experience: Comfort and convenience, considering factors like weight, size, wearing, and potential fatigue or discomfort.

\subsection{Methods}
\subsubsection{Participants}
Same as the fingertip pressure perception experiment(\ref{sec:presureExperiment}).
\subsubsection{System Set-Up and Experimental Procedure}

Regarding the hardware setup, we used an HMD due to the better experience it provides for dexterous manipulation in stereo vision. We utilized the Meta Quest Pro as the display hardware and attached a Leap Motion Controller (as it performed better than the built-in hand tracking), shown in Figure \ref{manipulationPhoto} b. Both hands were equipped with the proposed device.

Then, we conducted a variation of peg-in-hole task, which required the user to insert an object into the hole from bottom to top, as shown in Figure \ref{UnityScene} e. The object was placed on the table, measuring 3.5cm in length, 3.5cm in width, and 10cm in height. The setup included a platform with its lower surface 24 cm above the ground. This platform had a thickness of 2.5 cm and contained a through square hole, the hole on the platform has a square shape and the area is 3.95*3.95cm. A part of the platform is semi-transparent, allowing participants to see the inner surface of the hole, this design was made to reduce the difficulty of the experiment and save time.
Users had practice time and prompted with the phrase "Find a comfortable strategy to complete this task". The training concluded once the user thought they became familiar with the environment and could successfully complete the task. Upon the start of the formal experiment, users were instructed, "Use this strategy to complete the task 8 times." The experiment alternated between haptic feedback on and off, with the order being balanced. In "haptic on ", all of the feedback provided by the proposed device was turned on. If the object fell out of the scene, it would be reset to its initial position. We allowed users to intentionally do this, but the timing would not be reset. After the experiment, the participant was asked to fill out the interview questionnaire.

Questions:
\begin{itemize}
  \item What sensations or impressions do you have while using the device for grasping and manipulation?
  \item How do you think this method of operation is similar to or different from the experience of the real world?
  \item Do you think certain feedback modes (such as touch vibration, sliding vibration, etc.) have a positive or negative impact on your operational experience? Can you describe it in detail?

  \item Do you have comments on the wearing of the device, or what do you think the device can affect with your real-world behaviors?
  \item Do you have any other feelings or suggestions about your experience using the device?
\end{itemize}

\subsection{Results}
The peg-in-hole's result is shown in Figure \ref{peg}. We used the Wilcoxon signed-rank test to analyze the effect of conditions on the results(W=263.0, p\textless0.001), indicating a statistically significant difference between the haptics on and off. 
The statistical data for the "Haptic on" condition are as follows: median of 9.14 (SD=14.53, SEM=4.21). For the "haptic off" condition, the statistical data are as follows: median of 11.64 (SD=42.67, SEM=12.31).

\begin{figure}[!t]
\centering
\includegraphics[width=3.0in]{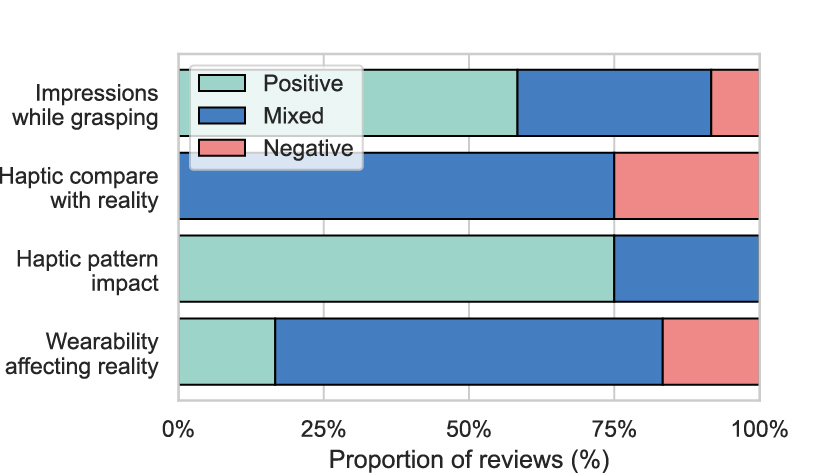}
\caption{A sentiment analysis of the interview responses, categorizing them as positive, mixed, and negative. The Mixed category includes answers that contain both positive and negative feedback. }
\label{interview}
\end{figure}

The sentiment analysis results from the interview are shown in Figure \ref{interview}. The majority of participants (9 out of 14) provided positive feedback on device usage, noting that vibration feedback improved grip control and realism in virtual object manipulation. One participant remarked, "I like the vibration when sliding; it helps adjust force," and another said, "The feel of touching the object aids in delicate operations." However, some participants (5 out of 14) found the haptic feedback limited in precise tasks, with comments like, "I can feel the touch, but it's debatable if it makes operations easier.";
Opinions on virtual versus real-world operations were mixed. Five participants felt virtual grasping was comparable to real touch pressure and position sensing. One noted, "Touch feels different, but with practice, it’s as good as real in sensing positions." Conversely, seven participants highlighted gaps in touch realism and contact area, with remarks like, "The operation demands more visual input than in the real world," and "The force and contact area are inconsistent.";
All participants had a non-negative attitude regarding the implemented haptic pattern. ”The touch vibration and sliding vibration do help my judgment on the operation”. But one participant thought the current feedback intensity wasn't sufficient to replace visual information;
Regarding wearability, six participants praised the device’s lightness and ease of wear. One participant noted that the cable arrangement didn't majorly affect operations. Conversely, six participants criticized issues like tangled cables and concerns about fully equipping the device. Safety concerns, such as potential burns from inductors, suggested a need for better heat isolation. Some participants also experienced a mismatch between visual and tactile sensations, affecting operations. Neutral feedback included suggestions for improved fixation methods and concerns about long-term use.

Participants also offered specific improvement suggestions, such as enhancing the fixation method, using area-based force feedback, and improving optical gesture recognition accuracy. One suggested, "I hope the force feedback generated by the strings could be changed to area-based feedback.

\subsection{Discussion}
In the peg-in-hole experiment, we identified several key factors. First, some users mentioned that "the lack of haptic feedback increases the difficulty of the operation" during the experiment. We observed that haptic feedback helps users avoid applying excessive or unbalanced forces when grasping objects, thereby reducing the probability of grasping failure. Second, during the process, when users attempt to gently insert a misaligned peg into the hole, although there is no vertical force feedback, the vibration cue indicates that the peg has collided with the outside of the hole, prompting users to adjust the position. Furthermore, once the peg enters the hole, force feedback assists users in precisely guiding the cube. We also noticed that users often use one hand to insert the object into the hole while gently pushing the object with the other hand. Haptic feedback allows users to accurately apply force to the object with their other hand without needing to shift their visual focus away from the hole, reducing operation time.

However, the accuracy of optical hand recognition (Leap Motion Controller) is insufficient for precise operations, leading to frequent recognition failures and errors, causing objects to accidentally drop and increasing the randomness of the results. Some users' gestures are difficult to recognize, resulting in significantly longer experiment times. 
Nevertheless, the experimental data indicate that the haptic feedback generated by our device improves the efficiency of dexterous manipulation. However, in the absence of kinesthetic constraints, haptic feedback has a limited effect on adjusting the object to align with the hole. Applying tangential forces with the fingers to determine the contact between the peg and the hole is still crucial.

In the interview section, despite omitting many haptic elements to keep the device lightweight, most participants found the haptic feedback acceptable and enhancing for virtual manipulation. This indicates that with the help of the physics engine, the haptic feedback from the device provides users with sensations that correspond to their interactions with these objects. Our haptic rendering also provided valuable information;
Another issue we are concerned about is the comparison with real-world operations, especially without applying kinesthetic constraints to the fingers. Some users have adapted to this, while others still feel it is not quite consistent with reality. Several participants found it difficult to perceive edges and surfaces, this suggests that displaying edges, surfaces, and hand resistance is still important. Furthermore, the participants generally considered the device's weight acceptable;
We are focusing on how our solution affects hand movements. Some participants appreciated the device's lightness and easy to wear, noting it doesn't hinder real-world tasks. Also there's some negative feedback like cable interference and uncomfortable heat. These problem can be addressed with engineering solutions such as wireless improvements and repositioning heating components. Additionally, some participants found the device difficult to wear due to reusable adhesive tape and the need to clean nails and match fingers, which also led to cable entanglement.

\section{Real-World Task Experiment}
To verify the advantages of the proposed devices not occupying the fingertips and not hindering real-world operations, we designed experiments that allowed users to perform daily tasks while wearing the device and compared their performance to bare hands and glove-type haptic devices. We selected representative tasks: typing and tightening a nut. We recorded the time participants took to complete these tasks and collected their subjective evaluations. 

\label{RealWorldExp}

\subsection{Methods}
\subsubsection{Participants}
Same as the fingertip pressure perception experiment(section \ref{sec:presureExperiment}).

\subsubsection{System Set-Up and Experimental procedure}

\begin{figure}[!t]
\centering
\includegraphics[width=2.7in]{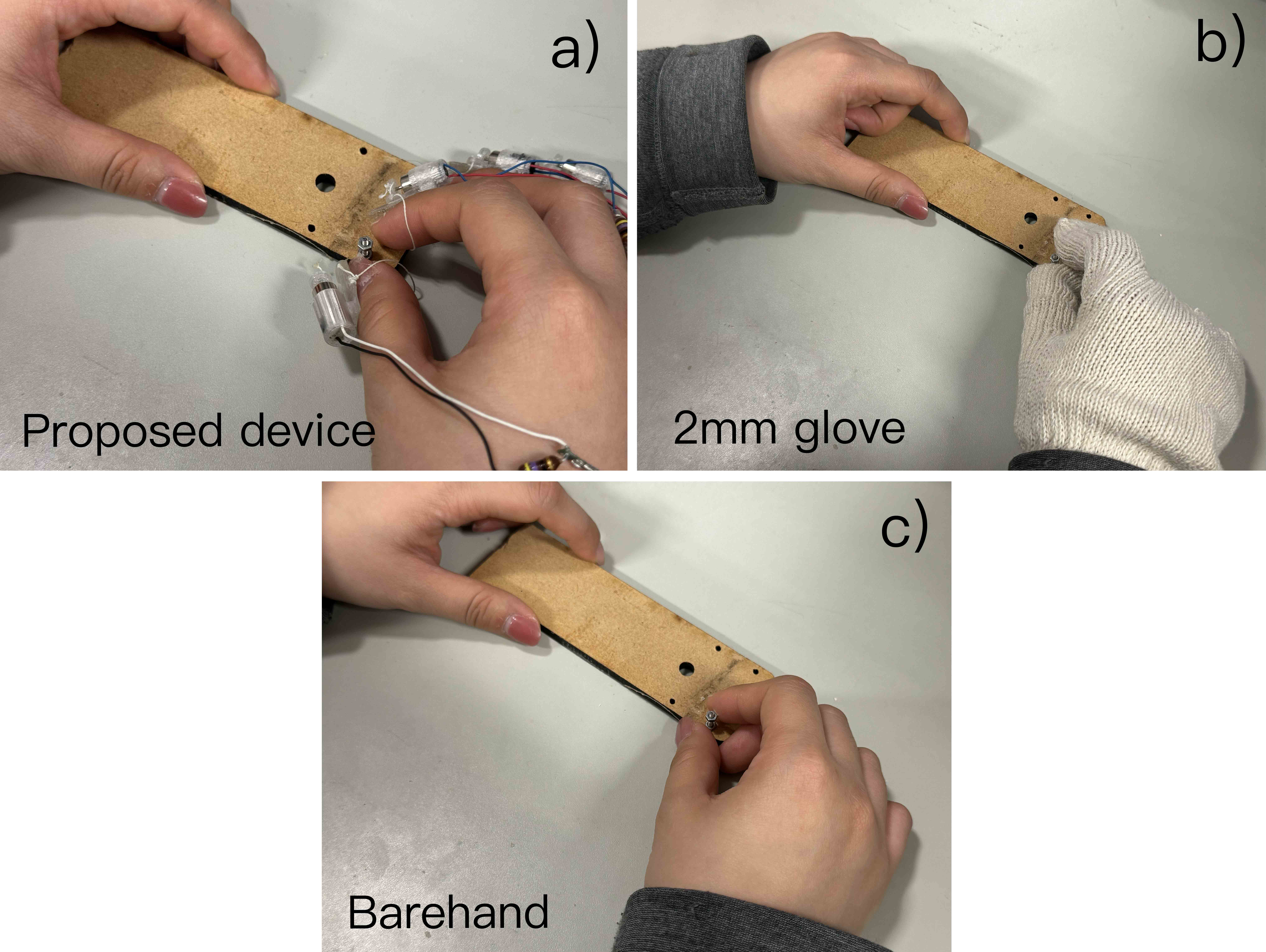}
\caption{The nut-tightening task was performed under three hardware conditions. The complete experimental process involved picking up a nut from the table and then attempting to nut it until it fully tightened. a) shows a participant wearing the proposed device, and b) shows a participant wearing a 2mm thick cloth glove to simulate other similar haptic devices. This experiment was designed to investigate the impact of haptic devices on manipulating real objects.}
\label{3types}
\end{figure}

In this experiment, we did not randomize the experimental order, allowing participants to transition directly between conditions and avoiding the inconvenience of repeatedly wearing and removing the device. The device setup is shown in Figure \ref{3types}. The experimental order was "Proposed device", "glove", and "barehand". Under each wearing condition, typing and nut-tightening experiments were conducted sequentially. To minimize the order effect, each participant was asked to perform pre-experimental training under all three conditions before formally recording the experimental results.  Moreover, the glove used in the experiment had a fabric thickness of 2mm.

During the typing task, participants sat in front of a desk, using a 13-inch Macbook Air M2 laptop with English input. The experiment included three typing tasks. Each task required the input of three different pangrams (a sentence containing every letter of the alphabet at least once) with 43 letters per sentence to reduce learning effects. Timing started with the entry of the first character and stopped upon completion of the last character.

The other experiment is tightening a nut. During the experiment, the screw was fixed on a baseboard, and participants could only use their right hand to turn the nut. Timing began when the participant's hand touched the nut and ended when the nut was fully tightened. The experiment sequence was: Typing Task 1 and nut-tightening task with the proposed device; Typing Task 2 and nut-tightening task with a glove; Typing Task 3 and nut-tightening task with bare hands.

After the experiment, participants were asked to fill out a questionnaire. The main question was "Regarding the experience of operating in the real world while wearing a haptic device, please score each experience." Participants were to give a score for each condition on a scale from 1 (very poor) to 7 (excellent).

\subsection{Results}

\begin{figure}[!t]
\centering
\includegraphics[width=3.0in]{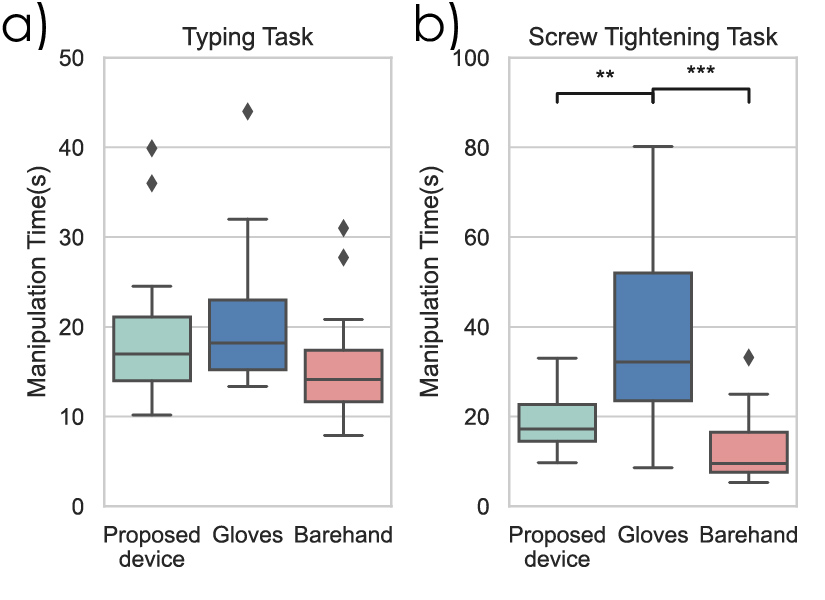}
\caption{User's manipulation time of typing and nut tightening tasks under three wearing conditions}
\label{manipulationTime}
\end{figure}

Manipulation task results are shown in Figure \ref{manipulationTime}. In the typing task, median values for "proposed device", "glove", and "barehand" were 16.99s (SD=9.308), 18.22s (SD=9.215), and 14.11s (SD=7.147). The Shapiro-Wilk test showed normal distribution for "barehand" (W=0.875, p\textgreater0.05), but not for "proposed device" (W=0.831, p\textless0.05) and "glove" (W=0.796, p\textless0.01). Levene's test indicated equal variances (W=0.113, p\textgreater0.05). The Kruskal-Wallis test showed no significant differences among conditions (H(2)=3.622, p=0.164).

In the nut-tightening task, median values for "proposed device", "glove", and "barehand" were 17.23s (SD=6.917), 32.13s (SD=23.907), and 9.54s (SD=8.787). The Shapiro-Wilk test showed normal distribution for "proposed device" (W=0.949, p\textgreater0.05) and "glove" (W=0.919, p\textgreater0.05), but not for "barehand" (W=0.821, p\textless0.05). Levene's test indicated unequal variances (W=5.696, p\textless0.01). The Kruskal-Wallis test showed significant differences among conditions (H(2)=12.452, p\textless0.01). Post-hoc Mann-Whitney U tests with Bonferroni correction revealed significant differences between "glove" and "barehand" (p\textless0.01, Hedges' g=-1.380) and between "proposed device" and "glove" (p\textless0.05, Hedges' g=1.128), but not between "proposed device" and "barehand" (p\textgreater0.05, Hedges' g=-0.632).

\begin{figure}[!t]
\centering
\includegraphics[width=2.0in]{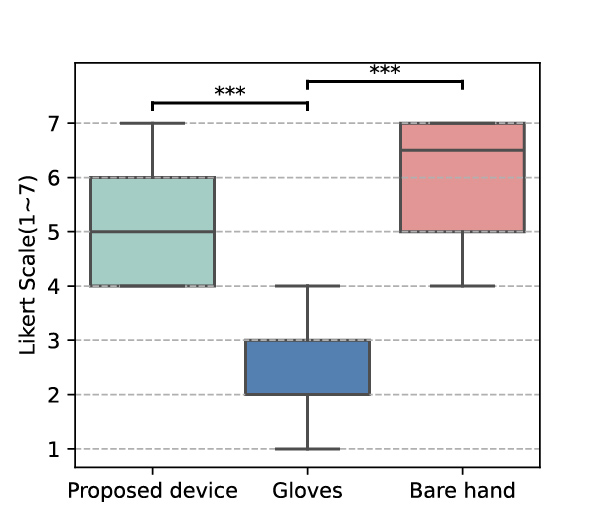}
\caption{Users’ ratings of manipulating experience under three wearing conditions}
\label{score}
\end{figure}

In the subjective evaluation shown in Figure \ref{score}, the Kruskal-Wallis test indicated significant differences among conditions (H(2)=24.929, p\textless0.001). Post-hoc Mann-Whitney U tests with Bonferroni correction revealed significant differences between "barehand" and "glove" (p\textless0.001) and between "proposed device" and "glove" (p\textless0.001), but not between "barehand" and "proposed device" (p\textgreater0.05).

\subsection{Discussion}
This experiment compared the performance of typing and nut-tightening tasks under three conditions (Proposed device, Glove, and Bare hand) and collected users' subjective ratings of their operational experience. In the typing tasks, statistical tests found no significant differences. This indicates that wearing gloves-like devices has a limited impact on typing speed, and users were able to maintain similar performance levels across various conditions.

However, in the nut-tightening task, the completion time while wearing the glove was significantly longer than the other two conditions, with statistically significant differences between the glove and both the Proposed device and Barehand. This suggests that gloves may reduce finger flexibility and tactile sensitivity, making precise operations more challenging. The performance of the proposed device was not significantly different from the bare-handed one, indicating that the device has limited restrictions on hand movements.

In the subjective rating, no significant difference was found between the "barehand" and "proposed device". This finding suggests that participants perceived the proposed haptic device as comparable to the barehand condition when performing the two tasks. Meanwhile, the glove condition may have impeded performance or comfort more than the other two conditions. These results also indicate that, compared to glove-type devices, the proposed device allows users to complete some daily tasks with less hindrance.

\section{Conclusions}
In this study, we developed a lightweight, wearable haptic device that provides physics-based haptic feedback for dexterous manipulation in virtual environments without hindering real-world interactions of fingertips. By employing thin strings and attaching actuators to the fingernails, the proposed device delivers pressure-based haptic experiences at the fingertips while maintaining the ability to interact with physical objects.

Through a series of experiments, we investigated the effectiveness of the proposed device in various scenarios. The fingertip pressure perception experiment demonstrated that participants could perceive and respond to the pressure feedback, the pinch distance can be larger than no haptic feedback when virtual object in hand. The slip vibrotactile feedback experiment showed that the vibration feedback improved their reaction time to the sliding task. The feedback can also significantly improve the efficiency of dexterous manipulation in the peg-in-hole experiment.

Furthermore, the real-world task experiments showed its ability to preserve natural tactile sensations and small hindrances to real-world operations. By keeping the palmar side of the fingers free, the device allows users to perform common tasks like typing and nut-tightening with similar performance to barehanded interactions.
The subjective evaluation provided insights into participants' experiences. Most participants appreciated the enhanced realism and precision provided by the haptic feedback, especially in judging the gripping state and adjusting the applied force. However, there were mixed opinions on the device's similarity to real-world interactions and its utility in precise operations. The device's wearability received both positive and negative feedback, with participants appreciating its lightness and ease of use. The proposed fingertip-based haptic devices enhance virtual manipulation experiences without compromising real-world interactions.

\bibliographystyle{ieeetr}
\bibliography{ref-base}

\vfill

\end{document}